\documentclass[showpacs,amssymb,aps]{revtex4}
\newcommand{\be}{\begin{equation}}
\newcommand{\ee}{\end{equation}}

\newcommand{\nnbb}{\nonumber\\}

\usepackage{epsf}
\usepackage{graphics}

\begin{document}

\title{Classical transport equation in non-commutative QED at high
temperature} 

\author{F. T. Brandt$^a$, Ashok Das$^b$, J. Frenkel$^a$}
\affiliation{$^a$ Instituto de F\'{\i}sica, Universidade de S\~ao
Paulo, S\~ao Paulo, SP 05315-970, BRAZIL}
\affiliation{$^b$Department of Physics and Astronomy, University
of Rochester, Rochester, NY 14627-0171, USA}

\bigskip

\bigskip

\date{\today}

\bigskip

\begin{abstract}
We show that the high temperature behavior of non-commutative QED may
be simply obtained from Boltzmann transport equations for classical
particles. The transport equation for the charge neutral particle is
shown to be characteristically different from that for the charged
particle. These equations correctly generate, for arbitrary values of
the non-commutative parameter $\theta$, the leading, gauge independent
hard thermal loops, arising from the fermion and the gauge sectors.  
We briefly discuss the generating functional of hard thermal amplitudes.
\end{abstract}


\maketitle

\section{Introduction}

The behavior of hot plasmas has been of considerable interest in
the recent years \cite{kapusta:book89lebellac:book96das:book97}. 
It is known, in particular, that in QCD the
leading behavior of the $n$-point gluon functions at temperatures
$T\gg p$, where $p$ represents a typical external momentum, is
proportional to $T^{2}$ and these leading contributions to the
amplitudes are all gauge independent \cite{Braaten:1990it,frenkel:1991ts}.
In order to extract the
leading order contributions to the amplitude leading to gauge
invariant results for physical quantities, it is necessary to
perform a resummation of hard thermal loops, which are defined by
\begin{equation}
p\ll k\sim T,
\end{equation}
where $k$ denotes a characteristic internal loop momentum. Such a
procedure, however, is quite
technical and an alternate simpler method, based on classical
transport equations, has been quite useful in deriving the gauge
invariant effective action which incorporates all the effects of
the hard thermal loops
\cite{Heinz:1985yq,Jackiw:1993zr,Nair:1993rx,Blaizot:1994be,
kelly:1994ig,Litim:sh}. 
In such an approach, one pictures the
constituents of the plasma as classical particles carrying color
charge and
interacting in a self-consistent manner. The main reason why
such an approach works is that, for soft gauge fields, the
occupation number per unit mode in a hot plasma is quite high due
to the Bose-Einstein enhancement.

More recently, following from developments in string theory, there
has been an increased interest in quantum field theories defined on
a non-commutative manifold satisfying
\cite{Seiberg:1999vs,Fischler:2000fv,Arcioni:1999hw,Landsteiner:2000bw,
Szabo:2001kg,Douglas:2001ba,Chu:2001fe,VanRaamsdonk:2001jd,Bonora:2000ga} 
\begin{equation}
[x^{\mu},x^{\nu}] = i \theta^{\mu\nu}
\end{equation}
where $\theta^{\mu\nu}$ is assumed to be a constant anti-symmetric
tensor with the dimensions of length squared. Furthermore, to
avoid problems  with unitarity, it is generally assumed that
$\theta^{0\mu}=0$ so that only the spatial coordinates are supposed
to have non-commutativity. We will assume this in our entire
discussion. The behavior of hot plasmas in such a non-commutative
gauge theory is an interesting question. Let us note that, in
non-commutative theories, because of the presence of a
dimensional parameter, the hard thermal loop approximation allows
for two extreme regions, namely,
\begin{equation}
\theta p T\gg 1,\qquad {\rm and}\qquad \theta p T\ll 1
\end{equation}
where perturbative calculations become simple. Here $\theta$ can
be thought of as the magnitude of $\theta^{\mu\nu}$. In
particular, the perturbative calculations \cite{Brandt:2002aa}
show that in the region
$\theta p T\gg 1$, all the oscillatory terms in the perturbative
amplitudes are negligible and the leading behavior is proportional to
$T^{2}$ as in ``conventional'' Yang-Mills theories (although
$\theta$ dependent). However, in the other  extreme limit, $\theta
p T\ll 1$, the leading behavior is suppressed and is proportional
to  $T^{2} (\theta pT)^{2}$.

In an earlier paper \cite{Brandt:2002rw}, 
we had proposed a classical transport
equation which reproduced the correct hard thermal loop amplitudes
in the region $\theta p T\gg 1$. In this regime, of course, one is
not expected to see the extended (dipole) nature of
non-commutative quanta \cite{Minwalla:1999px,Sheikh-Jabbari:1999vm,Bigatti:1999iz}
and a classical description based on the
picture of a plasma consisting of charged constituents 
suffices. However, away from this extreme region, the extended
nature of the non-commutative quanta will become relevant and will
play 
an important role. It is important, therefore, to ask if the
leading behavior of the amplitudes, at high temperature, can be
described by a classical transport equation for arbitrary values
of $\theta$. It is not obvious {\em a priori} that such a
description will work since it is known from the study of
``conventional'' QCD that the classical transport equation does not
yield the correct contributions for the hard thermal loop
amplitudes that are not proportional to $T^{2}$. In fact, such
(sub-leading) contributions  in ``conventional'' QCD are gauge
dependent while the transport equation is manifestly gauge
covariant. Let us note that, in general, the perturbative hard
thermal loop contributions can be written as $T^{2} H(\theta p T)$
where $H$ is a given function of the dimensionless variable
$\theta pT$. For $\theta pT\gg 1$, the leading behavior of
$H(\theta pT)\sim 1$ while for $\theta pT\ll 1$, the leading
behavior has the form $H(\theta pT)\sim (\theta pT)^{2}$. Although
for arbitrary values of $\theta$, the hard thermal loop amplitudes
cannot be evaluated in closed forms, nonetheless, it is easy to
check from their  integral representations that the leading terms
in $H(\theta p T)$ are gauge independent. Therefore, it would seem
that a classical description, based on a transport equation, may
reproduce the leading terms in the hard thermal loop approximation
for arbitrary $\theta$.

In this paper, we describe classical transport equations, for
QED, which precisely reproduce the leading hard thermal loop
contributions for arbitrary values of $\theta$. We show that the
transport equation for the charge neutral particle is
characteristically different from that for the charged particle. In
section {\bf II},  we give the results for the leading  hard thermal loop
amplitudes obtained from
perturbation calculations in an integral form, both for the
fermion as well as the photon and the ghost loops. In section {\bf III},
we describe the classical transport equation for a 
non-commutative charged particle, which reproduces all the hard
thermal loop amplitudes for the fermion loop for arbitrary
$\theta$. In section {\bf IV}, we describe the transport equation
for a charge neutral non-commutative particle which reproduces the
hard thermal loop amplitudes for the photon and the ghost
loops. In section {\bf V}, we derive from the transport equation a
manifestly covariantly conserved form of the current, which is related
to the generating functional of hard thermal loops.

\section{Results from perturbation theory}

Let us consider the Lagrangian density for non-commutative QED of
the form
\be\label{Lym}
{\cal L} = - \frac{1}{4} F_{\mu\nu}\star F^{\mu\nu} +  \bar{\psi}\star
(i\gamma^{\mu}D_{\mu}-m) \psi
\ee
where
\begin{equation}
D_{\mu}\psi = \partial_{\mu}\psi -
ie\left[A_{\mu},\psi\right]_{\rm MB}, \qquad F_{\mu\nu} =
\partial_{\mu}A_{\nu} - \partial_{\nu}A_{\mu} - ie
\left[A_{\mu},A_{\nu}\right]_{\rm MB}
\end{equation}
Here the Moyal bracket is defined to be
\begin{equation}
\left[A,B\right]_{\rm MB} = A\star B - B\star A
\end{equation}
where the Gr\"{o}newold-Moyal star product has the form
\begin{equation}
A(x)\star B(x) = \left.e^{\frac{i}{2}
\theta^{\mu\nu}\partial_{\mu}^{(\zeta)}\partial_{\nu}^{(\eta)}}
A(x+\zeta)B(x+\eta)\right|_{\zeta=0=\eta}
\end{equation}
We can add, to this Lagrangian density,  gauge fixing and ghost
terms which, in a covariant gauge, have the form
\begin{equation}\label{Lgf}
{\cal L}_{\rm gf} + {\cal L}_{\rm ghost} = -\frac{1}{2\xi}
(\partial_{\mu}A^{\mu})\star (\partial_{\nu}A^{\nu}) +
\partial^{\mu}\bar{c}\star (\partial_{\mu}c -
ie\left[A_{\mu},c\right]_{\rm MB})
\end{equation}
with $\xi$ representing the gauge fixing parameter.

\begin{figure}[h!]
\includegraphics*{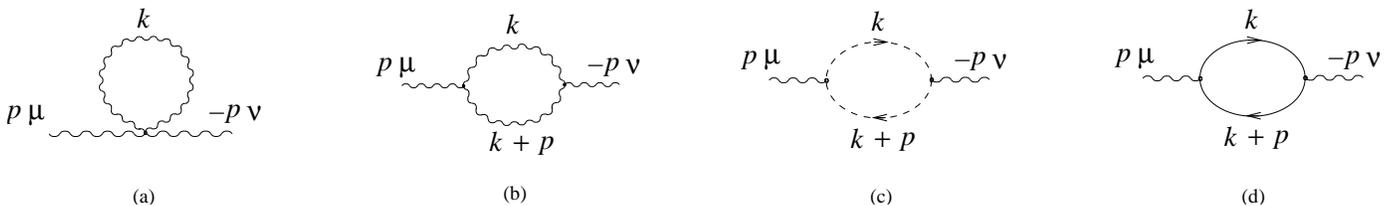}
\caption{One-loop diagrams which contribute to the self-energy
in the non-commutative QED. Wavy, dashed and solid lines denote respectively 
photons, ghosts and electrons. All the external momenta are incoming.}
\label{self}
\end{figure}

\begin{figure}[h!]
\includegraphics*{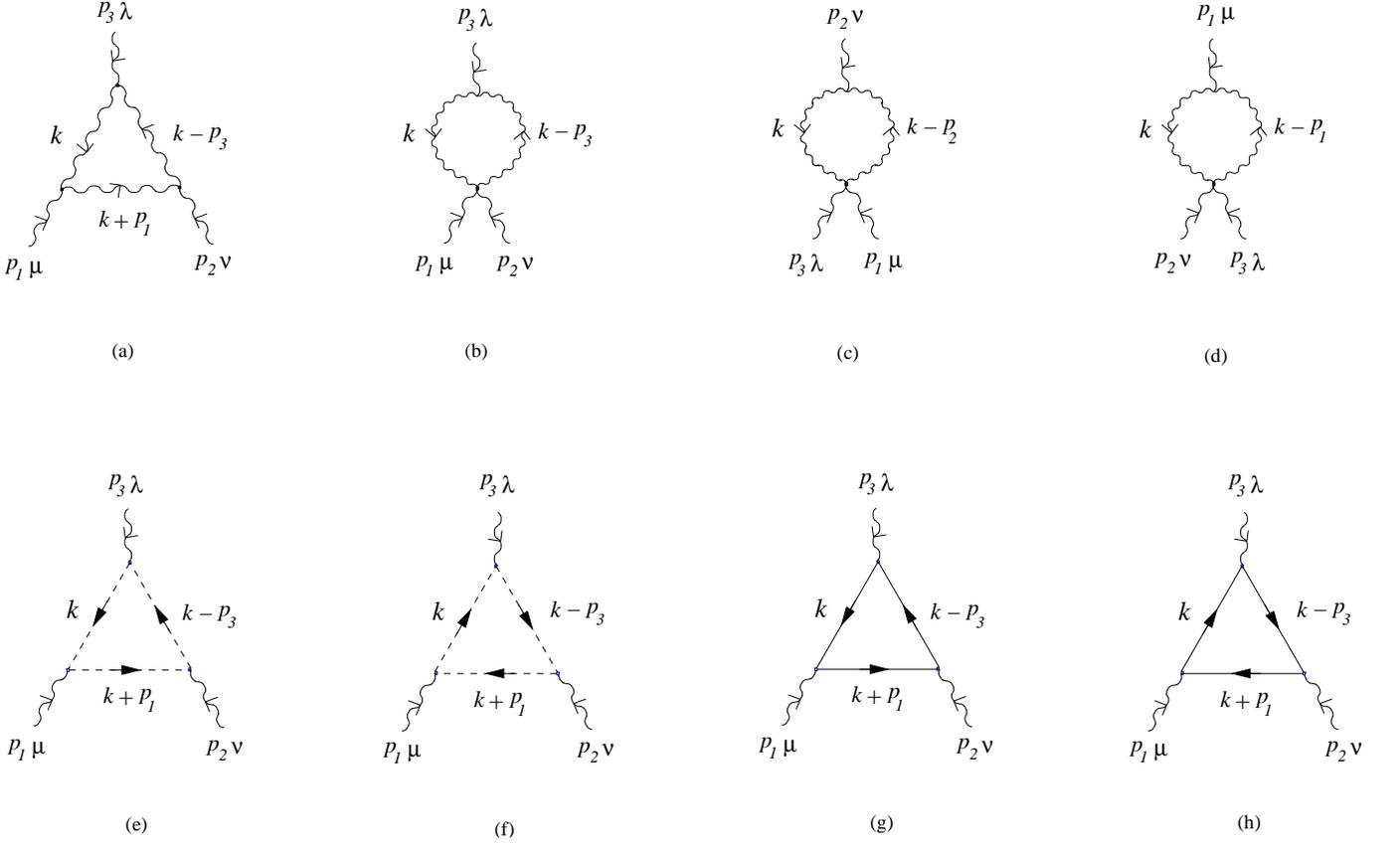}
\caption{One-loop diagrams which contribute to the three photon function
in the non-commutative QED. Wavy, dashed and solid lines denote respectively 
photons, ghosts and electrons. All the external momenta are incoming.}
\label{three}
\end{figure}

Our method of calculation employs an analytic continuation of the
imaginary-time formalism \cite{kapusta:book89lebellac:book96das:book97}.
Using this approach, we relate the Green's functions to forward scattering
amplitudes of on-shell thermal particles, a technique that has
been previously applied in the $SU(N)$ gauge theory as well as in
gravity \cite{brandt:1993mj,brandt:1993dkbrandt:1997se}. The
one-loop Feynman graphs which contribute to the photon self-energy are
shown in figure \ref{self}, while the one-loop diagrams associated
with the three-point function are represented in figure \ref{three}.

Using the Feynman rules given in the
appendix, the leading order contributions to the photon amplitudes in
the hard thermal loop approximation,
arising from the fermion loop, are obtained to be
\be
\Pi_{\mu\nu}^{\rm (fermion)} = -\frac{8\,e^2}{(2\pi)^3}\int
\frac{{\rm d}^3 k}{|\vec k|}\frac{1}{{\rm e}^{\frac{|\vec
k|}{T}}+1} \, \left.
G_{\mu\nu}(k;\,p)\right|_{k_0=|\vec k|}\label{fermionselfenergy} \ee
\be
\Gamma_{\mu\nu\lambda}^{\rm (fermion)} =
-\frac{8\,i\,e^3}{(2\pi)^3}\,\sin\left(\frac{p_1\times
p_2}{2}\right)\int \frac{{\rm d}^3 k}{|\vec k|}\frac{1}{{\rm
e}^{\frac{|\vec k|}{T}}+1} \, \left.L^{\rm
(fermion)}_{\mu\nu\lambda}(k;\,p_1,p_2,p_3)\right|_{k_0=|\vec k|},\label{fermionvertex}
\ee 
where the Lorentz structures are defined as
\be
G_{\mu\nu}(k;\,p)= \eta_{\mu\nu} - \frac{k_\mu\, p_\nu + k_\nu\,
p_\mu}{(k\cdot p)} + \frac{p^2\, k_\mu\, k_\nu}{(k\cdot p)^2}\label{G} \ee
\begin{eqnarray}
L^{\rm (fermion)}_{\mu\nu\lambda}(k;\,p_1,p_2,p_3) & = &
\left[\frac{p_1^2\,k_\mu\, k_\nu\, k_\lambda} {(k\cdot
p_1)^2(k\cdot p_3)}+ \frac{k_\mu\, k_\nu\, {p_3}_\lambda +
      k_\mu\, k_\lambda\, {p_3}_\nu}{(k\cdot p_1)(k\cdot p_3)}
+ \frac{k_\lambda}{(k\cdot p_3)} \eta_{\mu\nu} -
(\mu,p_1)\leftrightarrow (\lambda,p_3) \right]\nonumber\\ 
&  & \nonumber\\
&  &\qquad + {\rm
cyclic\; permut\; of} (\mu,p_1)\;(\nu,p_2)\;(\lambda,p_3)
\end{eqnarray}
Similarly, the leading order contributions to the photon
amplitudes coming from the photon and the ghost loops are determined
to be
\be
\Pi_{\mu\nu}^{\rm (gauge)} = -\frac{4\,e^2}{(2\pi)^3}\int
\frac{{\rm d}^3 k}{|\vec k|}\frac{1}{{\rm e}^{\frac{|\vec
k|}{T}}-1} \, \left(1-\cos(p\times k)\right)
\,\left.G_{\mu\nu}(k;\,p)\right|_{k_0=|\vec k|}\label{gaugeselfenergy}
\ee and
\begin{eqnarray}
\Gamma_{\mu\nu\lambda}^{\rm (gauge)} & = &
-\frac{4\,i\,e^3}{(2\pi)^3}\,\sin\left(\frac{p_1\times
p_2}{2}\right)\int \frac{{\rm d}^3 k}{|\vec k|}\frac{1}{{\rm
e}^{\frac{|\vec k|}{T}}-1} \left\{\left[1-\cos(p_3\times
k)\right]\, L^{\rm
(gauge)}_{\mu\nu\lambda}(k;\,p_1,p_2,p_3)\right. \nonumber\\ & & \qquad
\qquad \qquad \qquad \qquad \qquad \qquad \qquad\;\; -
\left[1-\cos(p_2\times k)\right]\, L^{\rm
(gauge)}_{\mu\lambda\nu}(k;\,p_1,p_3,p_2) \nonumber\\ & & \qquad \qquad
\qquad \qquad  \qquad \qquad \qquad \qquad\;\; \left. \left. -
\left[1-\cos(p_1\times k)\right]\, L^{\rm
(gauge)}_{\lambda\nu\mu}(k;\,p_3,p_2,p_1) \right\}\right|_{k_0=|\vec k|}
,\label{gaugevertex}
\end{eqnarray}
where
\begin{eqnarray}
L^{\rm (gauge)}_{\mu\nu\lambda}(k;\,p_1,p_2,p_3) & = & k_\mu\,
k_\nu\, k_\lambda \left(\frac{p_1^2}{(k\cdot p_1)^2(k\cdot p_3)}+
      \frac{p_1^2}{(k\cdot p_1)^2(k\cdot p_2)}-
      \frac{p_3^2}{(k\cdot p_3)^2(k\cdot p_1)}\right) \nonumber\\
& &\qquad + \frac{1}{(k\cdot p_2)(k\cdot p_3)} \left[ k_\nu\,
k_\lambda\left(p_2-p_3\right)_{\mu} + k_\mu\left(k_\lambda\,
{p_2}_\nu - k_\nu\, {p_3}_\lambda\right) \right] \nonumber\\
 & &\qquad +
\frac{1}{(k\cdot p_1)(k\cdot p_2)} k_\mu\left(k_\lambda\,
{p_2}_\nu + k_\nu\, {p_2}_\lambda\right) -2\frac{k_\mu}{(k\cdot
p_1)}\, \eta_{\nu\lambda}\nonumber\\
 & &\nonumber\\
 & &\qquad - (\mu,p_1)\leftrightarrow (\nu,p_2).
\end{eqnarray}

In computing these results, the gauge parameter $\xi$ in 
Eq. (\ref{Lgf}) has been kept arbitrary, but the dependence on $\xi$
cancels out in the final result in the leading order terms. The above hard
thermal loop amplitudes are gauge independent and satisfy simple Ward
identities. For example, from the structure in (\ref{G}), one can
easily  verify the transversality of the photon self-energy
\be
p^\mu\, \Pi_{\mu\nu}(p) = 0,
\ee
for both $\Pi_{\mu\nu}^{\rm (fermion)}$ and $\Pi_{\mu\nu}^{\rm
(gauge)}$. Similarly, the identity relating the two and the three
point functions 
\be
p_3^\lambda\,\Gamma_{\mu\nu\lambda}(p_1,p_2,p_3) = 2\,i\,e
\sin\left(\frac{p_1\times p_2}{2}\right)
\left[\Pi_{\mu\nu}(p_1)-\Pi_{\nu\mu}(p_2)\right].
\ee
can also be seen to hold separately for the contributions from the
fermion and the gauge sectors. This suggests the possibility that
these leading order hard thermal loop amplitudes may be obtained from
gauge covariant, classical transport equations.

\section{Transport equation for charged particles}

The basic idea behind the transport equation approach is to
picture the  thermal particles, moving in an internal loop, as
classical particles in equilibrium in the hot plasma whose
dynamics is governed by the classical transport equation. The
transport equation can, of course, be derived in a straightforward
manner once we know the dynamical equations for a particle in the
background of an electromagnetic field. Let us assume that the
equations of motion for a particle is given by
\begin{eqnarray}
m \frac{dx^{\mu}}{d\tau} & = & k^{\mu}\\ m \frac{dk^{\mu}}{d\tau}
& = & e X^{\mu}\label{force}
\end{eqnarray}
where $\tau$ denotes the proper time of the particle and $X^{\mu}$
represents  the force it feels in the presence of a background
electromagnetic field. The explicit form of $X^{\mu}$ will, of
course, be different depending on whether a particle is  charged
or neutral and we will discuss these two cases separately. In
general, the form of $X^{\mu}$ must be such that
$k^{2}=k^{\mu}k_{\mu}$ is a constant and that the time evolution
for $k^{\mu}$ transforms covariantly under a gauge transformation.
For the present, let us note that, given the equations for the
dynamics of the particle, the transport equation can be derived in
a straightforward manner as follows.

Let us define the current associated with the particle as
\begin{equation}
j_{\mu} (x) = e\,\sum\,\int dK\,k_{\mu} f(x,k)\label{current}
\end{equation}
where the sum is over helicities (One must also allow for a sum over
particle species in the case where there are more particle
types.). $f(x,p)$  represents the distribution function  for the
particle and the integration measure is defined to be
\begin{equation}
dK = \frac{d^{4}k}{(2\pi)^{3}}\,2\theta (k_{0})\,\delta
(k^{2}-m^{2})
\end{equation}
so as to  guarantee that the particle has positive energy and is
on-shell. This  is, of course, a natural generalization of the
conventional definition of the current to a non-commutative theory,
which will reduce naturally to the usual current in the
commutative limit. However, in the non-commutative theory, the
current transforms in the adjoint representation of the $U(1)$
group and correspondingly, the distribution function must also
transform covariantly under a $U(1)$ gauge transformation. The
current is covariantly conserved,
\begin{equation}
D_{\mu}j^{\mu} = \partial_{\mu}j^{\mu} - ie
\left[A_{\mu},j^{\mu}\right]_{\rm MB} = 0\label{conservation}
\end{equation}
as a consequence of the fact that it belongs to the adjoint
representation of the gauge group.

Given the equations for the particle and the covariant nature of
the distribution function, the transport equation has the general
form
\begin{equation}
D_{\tau} f (x,k) = {\cal C}
\end{equation}
where $D_{\tau}$ is the covariant derivative along the trajectory
and ${\cal C}$, in general, can be thought of as a collision term.
Using the equations of motion for the particle, we can rewrite
this as
\begin{equation}
k^{\mu} D_{\mu}f (x,k) + eX^{\mu}\star \frac{\partial f}{\partial
k^{\mu}} = m\, {\cal C}\label{transport1}
\end{equation}
It is a simple matter to check from the definition of the current
in (\ref{current}) and the transport equation in
(\ref{transport1}) that current conservation,
(\ref{conservation}), will hold provided we identify
\begin{equation}
{\cal C} = - \frac{e}{m} \frac{\partial X^{\mu}}{\partial
k^{\mu}}\star f(x,k)
\end{equation}
so that we can write the transport equation, in general, as
\begin{equation}
k^{\mu} D_{\mu} f(x,k) + \frac{\partial (e X^{\mu}\star
f(x,k))}{\partial k^{\mu}} = 0\label{transport}
\end{equation}

We note here that, for a ``conventional'' charged point particle, we
have
\begin{equation}
X^{\mu}_{\rm (conventional)} = F^{\mu\nu}k_{\nu}
\end{equation}
so that
\begin{equation}
{\cal C}_{\rm (conventional)} = 0
\end{equation}
Therefore, in such a case, one can naturally talk of a
``collisionless" plasma. However, we note that, in principle, if
$X^{\mu}$ is a more complicated function of $k^{\mu}$, then, a
basic ``collision" term is inevitable in discussing a hot plasma.
This will become quite clear when we discuss the transport
equation for a charge neutral particle in the next section.

The particles, in a non-commutative theory, have dipole
characteristics. Therefore, when we talk of a charged particle in
such a theory, it is natural to generalize the form of the force
to include a dipole term, namely,
\begin{equation}
X^{\mu}_{\rm (charged)} = \left(1 + \sin {k\times iD}\right)
F^{\mu\nu} k_{\nu}\label{chargeforce}
\end{equation}
where we have introduced, for simplicity, the standard notation
(within the context of non-commutative theories) that
\[
A\times B = \theta^{\mu\nu} A_{\mu}B_{\nu}
\]
and $D_{\mu}$ is the covariant derivative introduced earlier in
the adjoint representation. The first term, on the right hand side
of (\ref{chargeforce}), represents the standard Lorentz force for a
point charged particle. The second term, when the external momentum
is small (or the variations of the external fields are slow), can be
thought of as a dipole interaction, which would be a natural
generalization considering that non-commutative particles have
dipole characteristics. Furthermore, this term vanishes when
$\theta^{\mu\nu}\rightarrow 0$ so that the force reduces naturally
to the conventional one for a charged point particle. If such an
additional term is present in the equation of motion
(\ref{force}), it is clear that we will have a non-trivial
contribution to the transport equation coming from ${\cal C}$.
However, {\em a posteriori}, it turns out that such a dipole
interaction term 
is not manifest in the leading perturbative amplitudes
at high temperatures. Namely, the presence of such a dipole
interaction term is incompatible with the required symmetries of the
two and three point photon amplitudes in the leading order. This is
also  otherwise
clear, namely, we know that fermion loops (charged loops) only
give planar (but $\theta$ dependent) contributions.

With this, we conclude that the correct transport equation for a
charged particle, in the presence of (non-commutative) background
electromagnetic fields, has the form
\begin{equation}
k^{\mu}D_{\mu} f(x,k) + \frac{\partial (ek_{\nu} F^{\mu\nu}\star
f(x,k))}{\partial k^{\mu}} = 0\label{charged}
\end{equation}
where, for hard thermal loop contributions, we have effectively
\begin{equation}
X^{\mu}_{\rm eff} = F^{\mu\nu} k_{\nu}
\end{equation}
Equation (\ref{charged}) is manifestly gauge covariant and, although
its  form
is very similar to the transport equation for ``conventional'' QED,
the star products bring in the necessary $\theta$ dependence into
the amplitudes.

Let us recall that the distribution function is determined from
the transport equation iteratively in order to obtain the current
(\ref{current}), which yields the amplitudes. Thus, writing
\begin{equation}
f(x,k) = f^{(0)} + e f^{(1)} + e^{2} f^{(2)} + \cdots
\label{expansion}
\end{equation}
and substituting this into the transport equation (\ref{charged}), we
obtain,  to
the lowest order,
\begin{equation}
f^{(0)} \sim n_{B}(|k^{0}|)\qquad {\rm or}\qquad n_{F}(|k^{0}|)
\end{equation}
which are the conventional equilibrium distribution functions for
bosons and fermions, depending on the nature of the charged
particle in the loop. In the present case, where we are considering
the contributions from the fermion loop, we choose $f^{(0)} \sim
n_{F}(|k^{0}|)$.  With this, we see that, to the lowest order in $e$,
the current (\ref{current}) vanishes (actually, the time component
of the current to the lowest order is a constant which can be set
to zero by a simple redefinition). Substituting the lowest order
solution, $f^{(0)}$, into the transport equation (\ref{charged}),
leads to
\begin{equation}
f^{(1)} = \frac{1}{k\cdot \partial}\, \frac{\partial}{\partial
k_{\nu}} \left((k\cdot \partial A_{\nu} - \partial_{\nu} k\cdot
A)f^{(0)}\right)
\end{equation}
Substituting this into the definition of the current,
(\ref{current}), we obtain the leading contribution of the
fermions to the photon self-energy as
\begin{equation}
\Pi_{\mu\nu}^{\rm (fermion)} (p) = \left.\frac{\delta j_{\mu}
(p)}{\delta A^{\nu} (-p)}\right|_{A_{\mu}=0} =  -
\frac{8\,e^{2}}{(2\pi)^{3}} \int \frac{d^{3}k}{|\vec{k}|}\, n_{F}
(|\vec{k}|)\,G_{\mu\nu} ({k};p).
\end{equation}
Here, and in the following, the on-shell condition
$k_0=|\vec k|$ is to be always understood. It is clear that this agrees
completely  with (\ref{fermionselfenergy}).

With the form of $f^{(1)}$ determined, we can iterate the
transport equation (\ref{charged}) to determine the distribution
function in the next order which has the form
\begin{equation}
f^{(2)} = \frac{i}{k\cdot \partial}\left(\left[k\cdot
A,f^{(1)}\right]_{\rm MB} - \left[k\cdot A,A_{\nu}\right]_{\rm MB}
\frac{\partial f^{(0)}}{\partial k_{\nu}} + \cdots\right)
\end{equation}
where ``$\cdots$'' represent terms that do not contribute at the
leading order to the current (and, therefore, to the three point
amplitude). Substituting this into the current, (\ref{current}),
we obtain the leading contribution of the fermions to the three
point photon amplitude as ($p_{1}+p_{2}+p_{3}=0$)
\begin{eqnarray}
\Gamma_{\mu\nu\lambda}^{\rm (fermion)} (p_{1},p_{2},p_{3}) =
\left.\frac{\delta^{2}j_{\mu}(p_{1})}{\delta A^{\nu}(p_{2})\delta
A^{\lambda}(p_{3})}\right|_{A_{\mu}=0} =
\frac{16\,i\,e^3}{(2\pi)^3}\,
\sin\left(\frac{p_1\times p_2}{2}\right)
\int\frac{{\rm d}^3\, k}{|\vec k|} n_F(|\vec k|)\frac{1}{k\cdot p_3}\nnbb
\left[G_{\mu\nu}(k; p_2)\, k_\lambda + \frac{k_\mu\, k_\nu}{k\cdot p_2}
\left(\frac{p_2\cdot p_3}{k\cdot p_3}\, k_\lambda - {p_2}_\lambda\right)
- (p_1\leftrightarrow p_2)\right].
\end{eqnarray}
This can be checked to be completely equivalent to the result from
perturbative calculations in (\ref{fermionvertex}).

\section{Transport equation for charge neutral particles}

The case of the transport equation for a charge neutral particle
is even more interesting and challenging. A ``conventional" charge
neutral  point
particle, of course, does not feel any electromagnetic force.
However, the particles in a non-commutative theory have an
extended structure (because of the non-commutativity of
coordinates). Consequently, even a charge neutral particle in a
non-commutative theory can have a dipole structure, as is
generally believed, and such a particle can feel a dipole force in
the presence of background electromagnetic fields. On the other
hand, such a dipole force, as we have argued in the last section,
is not manifest in the leading order amplitudes calculated from
perturbation theory. Therefore, one has to think more carefully
with the only guidance coming  from the explicit results of perturbation
theory. In principle, since the dipole force does not work, one
may try a properly covariantized quadrupole form of the
interaction in (\ref{force}) for a charge neutral particle in a
non-commutative theory. While this does satisfy all the necessary
requirements of gauge covariance, it leads to amplitudes which do
not quite agree with the actual perturbative results. This,
therefore, necessitates a systematic search for a modification of the
force beyond the quadrupole interaction.

With some detailed analysis, we find that, for a charge neutral
particle to  have a transport
equation which reproduces the hard thermal loop amplitudes coming
from the photon and the ghost loops, we must have
\begin{equation}
X^{\mu}_{\rm (charge\;neutral)} = 
2\left\{1-\cos k\times\left(i\, D + e\left[\frac{1}{k\cdot D}\, F,
\;\;\;\;\right]_{MB}\right)\right\}
\, F^\mu \label{neutralforce}
\end{equation}
Here, we have defined a useful (suggestive) vector
\begin{equation}\label{39}
F^{\mu} = F^{\mu\nu}k_{\nu}
\end{equation}
The meaning of this force will be discussed later. For the
present, let us note that with this form of $X^{\mu}$, equation
(\ref{force}) is manifestly gauge covariant and leads to $k^{2} =
{\rm constant}$. Furthermore, with this complicated form of the
force on a charge neutral particle, we note that the ``collision''
term can no longer be neglected. This is consistent with the
extended nature of particles in a non-commutative theory.

With the form of $X^{\mu}_{\rm (charge\;neutral)}$ in
(\ref{neutralforce}),  we can solve
for the distribution function iteratively from the transport
equation (\ref{transport}). The lowest order term is determined to
be proportional to the equilibrium boson distribution function,
\begin{equation}
f^{(0)} \sim n_{B}(|k^{0}|)
\end{equation}
which leads, as in the charged case, to a vanishing current in the
lowest order. At the next order, the distribution function is then
determined to be
\begin{equation}
f^{(1)} = \frac{2}{k\cdot \partial} \frac{\partial}{\partial
k_{\mu}} \left((1 - \cos k\times i\partial) (k\cdot \partial
A_{\mu} - \partial_{\mu} k\cdot A) f^{(0)} (k)\right)
\end{equation}
The current, to this order, can now be obtained and leads to the
photon two point function
\begin{equation}
\Pi_{\mu\nu}^{\rm (gauge)} (p) = \left.\frac{\delta j_{\mu}(p)}{\delta
A^{\nu}(-p)}\right|_{A_{\mu}=0}  =
- \frac{4\,e^{2}}{(2\pi)^{3}} \int
\frac{d^{3}k}{|\vec{k}|}\;n_{B}(|\vec{k}|)\, G_{\mu\nu} ({k};p)
(1- \cos p\times k)
\end{equation}
This, in fact, agrees completely with eq. (\ref{gaugeselfenergy})
and is purely transverse.

We can now use $f^{(1)}$ in the transport equation
(\ref{transport}) to determine the distribution function at the
next order. With a little bit of work, it is easy to see that
\begin{equation}
f^{(2)}  = \frac{1}{k\cdot \partial} \left(i\left[k\cdot A,
f^{(1)}\right]_{\rm MB} - \left.\frac{1}{e} \frac{\partial\left(
X^{\mu}_{\rm (charge\;neutral)}\star f\right)}{\partial
k^{\mu}}\right|_{\rm quadratic}\right)
\end{equation}
Here, the restriction ``quadratic" refers to quadratic terms in
the $A_{\mu}$ fields (or linear in $e$). Let us note that
$X^{\mu}_{\rm (charge\;neutral)}$ involves trigonometric functions
of non-commuting operators and the expansion of these functions,
even to linear order in $A_{\mu}$ (the trigonometric functions are
multiplied by a field strength), is highly nontrivial in the
coordinate space. Namely, one must use the
Baker-Campbell-Hausdorff formula for such an expansion which
involves an infinite series and {\em a priori}, it is not clear if
a closed form expression necessary for the computations (and
comparison with the perturbative results) can be obtained. The
solution lies in the fact that these operators take a much simpler
form in the momentum space. Namely, substituting this into 
(\ref{current}), the form of the current to second order
is determined to have a closed form expression in the momentum space given by
\begin{eqnarray}
j_{\mu}^{(2)}(p_{1}) & = & 8ie^{3} \int{\rm d}K {\rm d}^{4}p\,\sin 
\frac{p_{1}\times p}{2}\,
\frac{k_{\mu}}{k\cdot p_{1}}\left[\frac{k\cdot A(-(p_{1}+p))}{k\cdot p}
\frac{\partial((1-\cos k\times (p_{1}+p))H_{\sigma}(p,k) f^{(0)})}
{\partial k_\sigma}\right.\nnbb 
&  & \qquad\qquad - \frac{\partial}{\partial k_{\sigma}} 
\left((1-\cos (k\times p_{1})) k\cdot A(-(p_{1}+p)) A_{\sigma}(p)
f^{(0)}\right) \nnbb
 &  & \qquad \left. +  \frac{\partial}{\partial k_{\sigma}}
 \left(\frac{k\cdot A(-(p_{1}+p))}{k\cdot (p_{1}+p)} 
(\cos k\times p_{1} - \cos k\times p)H_{\sigma}(p,k) f^{(0)}\right)
+ \cdots\right]
 \end{eqnarray}
 Here, we have defined, for simplicity,
 \begin{equation}
 H_{\sigma}(p,k) = p_{\sigma} k\cdot A(p) - k\cdot p A_{\sigma}
 (p)
 \end{equation}
 and the ``$\cdots$" represent terms that do not contribute at the
 leading order.

 It is now easy to obtain the three point amplitude from the
 second order current ($p_{1}+p_{2}+p_{3}=0$)
 \begin{eqnarray}
 \Gamma_{\mu\nu\lambda}^{\rm (gauge)}(p_{1},p_{2},p_{3}) & = & 
\left.\frac{\delta^{2} j_{\mu}(p_{1})}{\delta A^{\nu}(p_{2})\delta
 A^{\lambda}(p_{3})}\right|_{A_{\mu}=0}\nnbb 
  & = & 
\frac{8\,i\,e^3}{(2\pi)^3}\,\sin\left(\frac{p_1\times p_2}{2}\right)
\int\frac{{\rm d}^3 k}{|\vec k|}\, n_B(|\vec k|)\frac{1}{k\cdot p_1}
\left\{\left[1-\cos(k\times p_1)\right]\,k_\lambda\,G_{\mu\nu}(k;p_1)
{{~}\atop{~}}\right.\nnbb
& & \qquad + \left[1-\cos(k\times p_3)\right]
\left[k_\mu\,G_{\nu\lambda}(k;p_3) + 
k_\nu\,G_{\mu\sigma}(k;p_1)\,G^\sigma_\lambda(k;p_3) \right] \nnbb
& & \quad + 
\left[\cos(k\times p_1)-\cos(k\times p_3)\right]
\left.\frac{k\cdot p_3}{k\cdot p_2}\,
k_\nu\,G_{\mu\sigma}(k;p_1)\,G^\sigma_\lambda(k;p_3)-
(p_2\leftrightarrow p_3; \nu\leftrightarrow \lambda)\right\}.
\end{eqnarray}
With the help of some simple algebraic identities, it may be verified
that this expression is in complete agreement with the diagrammatic result
(\ref{gaugevertex}). This is indeed a nontrivial check that the
transport equation for the charge neutral particle is, in fact,
correct.

\section{Summary and discussions}

In this paper, we have given  transport equations, for charged as
well as charge neutral particles, that reproduce exactly the leading two
and the three point photon amplitudes in non-commutative QED, in
the hard thermal loop approximation, 
{\em for arbitrary value of $\theta$}. Since the leading amplitudes 
satisfy tree level Ward identities
and the transport equation is manifestly gauge covariant, it
follows that these are the true equations which will generate all the
amplitudes in the leading hard thermal loop approximation, both for the
charged  as well as the charge neutral loops.

The force for the charge neutral particle, (\ref{neutralforce}), is
completely new and 
is worth discussing a little since it adds to the physical picture
of particles in non-commutative theories. We note that, in the
limit of small external momenta (or small variations in the
background), the leading order term in the force $X^{\mu}_{\rm
(charge\;neutral)}$ is that of a quadrupole (in fact, in the
configuration of a pair of dipoles aligned back to back). This is
quite interesting in the sense that while the conventional picture
of a charge neutral particle in a non-commutative theory is that
of a dipole, we see that the dipole interaction is not manifest in
the hard thermal loop approximation. Instead, the leading effect
comes from a quadrupole interaction. Furthermore, a local
quadrupole interaction (properly covariantized) is not sufficient,
rather the true equations need a further modification by a
nonlocal interaction.

Let us also note from the covariant form of the transport equation, 
(\ref{transport}), that we can write
\begin{equation}
f = - \frac{e}{k\cdot D} \frac{\partial (X^{\mu}\star f)}{\partial
  k^{\mu}} \label{integral}
\end{equation}
which provides an integral equation representation for the
distribution function. Substituting this into the definition of the
current, (\ref{current}), we obtain 
\begin{equation}
j_{\mu} = -2\,e^{2}\int dK\,\frac{k_{\mu}}{k\cdot D}\,
\frac{\partial (X^{\nu}\star f)}{\partial k^{\nu}}\label{current1}
\end{equation}
We note that this form of the current has several interesting
features. It transforms covariantly and is manifestly covariantly
conserved. Furthermore, even though equation (\ref{integral}) can only
be solved iteratively leading to an iterative solution for the current
(\ref{current1}), experience shows that the leading contributions to
the current, in the hard thermal loop approximation, can be obtained
from (\ref{current1}) by simply replacing $f\rightarrow f^{(0)}$ on
the right hand side. With
this, the  leading form of the current becomes
\begin{equation}
j_{\mu} = 2\,e^{2}\int dK\, \left(\eta_{\mu\nu} -
k_{\mu}\frac{1}{k\cdot D} D_{\nu}\right)\frac{1}{k\cdot D}\,X^{\nu} f^{(0)}
\end{equation}
Substituting the form of $X^{\mu}_{\rm (charge\;neutral)}$ given by Eq.
(\ref{neutralforce}) this takes the form
\begin{equation}
j_{\mu}^{\rm (charge\;neutral)} = \frac{4e^{2}}{(2\pi)^{3}} \int
\frac{d^{3}k}{|\vec{k}|}\,n_{B}(|\vec{k}|) \left(\eta_{\mu\nu} - k_{\mu}
\frac{1}{k\cdot D} D_{\nu}\right)\frac{1}{k\cdot D} 
\left\{1-\cos k\times\left(i\, D + e\left[\frac{1}{k\cdot D}\, F,
\;\;\;\right]_{MB}\right)\right\} F^{\nu}, \label{neutralcurrent}
\end{equation}
with $F$ defined in Eq. (\ref{39}).
We note that in the regime $k\times i\partial \gg 1$, where the
cosine oscillates rapidly and becomes
negligible, this reduces to our earlier result
\cite{Brandt:2002rw} (which shows only
planar contributions). In the commutative limit,
$\theta^{\mu\nu}\rightarrow 0$, the current vanishes as should be the
case for a charge neutral particle in a commutative
theory. Furthermore, it is worth noting here that, for a charged
particle, the current in the commutative limit, reduces to that in 
``conventional'' QED.

If we think of the current as arising from an effective action as,
\begin{equation}
j_{\mu}[A] = \frac{\delta\Gamma[A]}{\delta A^{\mu}}
\end{equation}
one can, in principle, functionally integrate (\ref{neutralcurrent})
to obtain the effective action which generates the hard thermal loops. 
However, except for the planar regime
\cite{Brandt:2002rw}, this is very difficult to carry out, in general.

The covariant conservation of the current (\ref{conservation}), ensures
that the effective action is gauge invariant. This follows because,
under a gauge transformation generated by the infinitesimal gauge
parameter $\omega(x)$,
\be
\frac{\delta\Gamma[A]}{\delta\omega(x)} = \int {\rm d}^4 y
\frac{\delta A_\mu(y)}{\delta\omega(x)}\,
\frac{\delta\Gamma[A]}{\delta A_\mu(y)} = - D_\mu\, j^\mu(x) = 0.
\ee
One may take advantage of the fact that the effective action is gauge
invariant, and try to evaluate it in a light-like axial gauge in which
$k\cdot A^{(k)} = 0$, with $k_0=|\vec k|$. In this gauge, it is possible to
integrate (\ref{neutralcurrent}), to obtain an effective action of the form
\be
\Gamma^{{\rm (charge\; neutral)}}[A] 
= - \frac{2\, e^2}{(2\pi)^3}\int\frac{{\rm d}^3 k}{|\vec k|}
n_B(|\vec k|)\int{\rm d}^4 x\, A^{(k)}_\mu(x)\star
\left[1-\cos(k\times i\partial)\right]\, {A^\mu}^{(k)}(x).
\ee
Although this expression has a deceptively simple structure, 
in reality, the $k$-integration is nontrivial due to the complicated
$k$-dependence implicit in the potential $A_\mu^{(k)}(x)$. This
question clearly deserves further study. 

\begin{acknowledgments}
We would like to thank Professor J.~C.~Taylor for many helpful discussions.
This work was supported in part by US DOE Grant number DE-FG
02-91ER40685 and by CNPq and FAPESP, Brazil.
\end{acknowledgments}

\appendix

\section{Feynman rules}

The Feynman rules following from the Lagrangian
given by Eqs. (\ref{Lym}) and (\ref{Lgf}) take the forms

\begin{eqnarray}
\includegraphics*{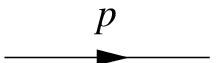} 
& : \;\;\; & \;\;\;
\displaystyle{{i\over p\!\!\!\slash - m + i\epsilon} =
iS(p)}\nonumber\\
\includegraphics*{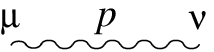}
& : \;\;\;&\;\;\; -\displaystyle{ {i\,\over (p^{2}+i\epsilon)}
\left(\eta_{\mu\nu} -
 (1-\xi){p_{\mu}p_{\nu}\over p^{2}}\right)} = i\, D_{\mu\nu}(p) \nonumber\\
\includegraphics{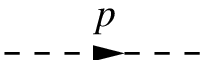}
 & : \;\;\;&\;\;\;
\displaystyle{{i\,\over p^{2} + i\epsilon}} = i\, D(p)
\label{apa1}\end{eqnarray}

\begin{eqnarray}
\begin{array}{c}
\includegraphics*{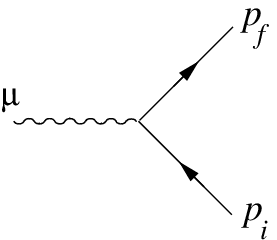}
\end{array}
 & :\;\;\;  & \;\;\;
ie\gamma^{\mu} e^{{i\over 2} p_{i}\times p_{f}}\nonumber\\ 
 & & \nonumber\\
\begin{array}{c}\includegraphics*{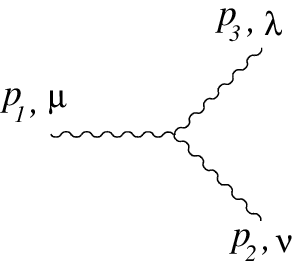}\end{array}
 & :\;\;\;  &\;\;\;  -2\,e\,\sin\left(\frac{p_1\times p_2}{2}\right)
\left[(p_{1}-p_{2})^{\lambda}\eta^{\mu\nu} +
      (p_{2}-p_{3})^{\mu}\eta^{\nu\lambda} +
      (p_{3}-p_{1})^{\nu}\eta^{\lambda\mu}\right]\nonumber\\
  & & \nonumber\\
\begin{array}{c}\includegraphics*{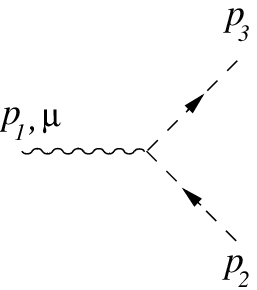}\end{array}
 & :\;\;\;  &\;\;\;  -2\,e\,\sin\left(\frac{p_2\times p_3}{2}\right)
 p_{3}^{\mu} \nonumber\\
 & & \nonumber\\
\begin{array}{c}\includegraphics*{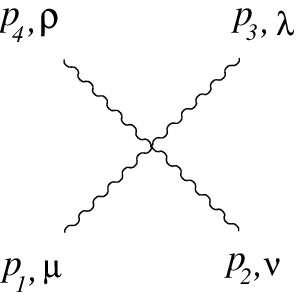}\end{array}
 & :\;\;\;  &\;\;\;  -4\,i\, e^{2}\left[
\sin\left(\frac{p_1\times p_2}{2}\right)\sin\left(\frac{p_3\times
p_4}{2}\right) (\eta^{\mu\lambda}\eta^{\nu\rho} -
\eta^{\mu\rho}\eta^{\nu\lambda}) \right.\nonumber\\
 &  & \qquad\quad +
\sin\left(\frac{p_1\times p_3}{2}\right)\sin\left(\frac{p_4\times
p_2}{2}\right) (\eta^{\mu\rho}\eta^{\lambda\nu} -
\eta^{\mu\nu}\eta^{\lambda\rho}) \nonumber\\ 
& & \nonumber\\ 
& &\nonumber\\
 &  & \qquad\quad\left. +
\sin\left(\frac{p_1\times p_4}{2}\right)\sin\left(\frac{p_2\times
p_3}{2}\right) (\eta^{\mu\nu}\eta^{\lambda\rho} -
\eta^{\mu\lambda}\eta^{\nu\rho}) \right],\nonumber\\ 
& & \label{vertex}
\end{eqnarray}
where {\it all momenta} are incoming and the Dirac delta functions
representing the conservation of momenta are understood.


\end{document}